\documentclass[letterpaper,prl,twocolumn,superscriptaddress,preprintnumbers]{revtex4}

\usepackage{graphicx}% Include figure files
\usepackage{dcolumn}% Align table columns on decimal point
\usepackage{bm}% bold math
\usepackage{wasysym}

\def\beq{\begin{equation}}
\def\eeq{\end{equation}}
\def\beqn{\begin{eqnarray}}
\def\eeqn{\end{eqnarray}}

\def\be{\begin{equation}}
\def\ee{\end{equation}}
\def\ba#1{\begin{array}{#1}}
\def\ea{\end{array}}
\def\bn{\begin{enumerate}}
\def\en{\end{enumerate}}
\def\r{\right}
\def\l{\left}
\def\H{\mathcal{H}}
\def\summ{\sum\limits}
\def\intt{\int\limits}

\def\o{\omega}
\def\x{\vec{x}}

\newcommand{\mysw}[1]{\scriptscriptstyle #1}

\begin{document}
%\preprint{\sf Version 6 (\today)}

\title{Ground-State Degeneracy of Correlated Insulators with Edges}
\author{Gil Refael}
\affiliation{Kavli Institute of Theoretical Physics, University of
  California, Santa Barbara, CA 93106}
\author{Hsiu-Hau Lin}
\affiliation{Kavli Institute of Theoretical Physics, University of
  California, Santa Barbara, CA 93106}
\affiliation{Department of Physics, National Tsing-Hua University,
  Hsinchu 300, Taiwan}
\affiliation{Physics Division, National Center for Theoretical Sciences, Hsinchu 300, Taiwan}
%\author{Masaki Oshikawa}
%\affiliation{Department of Physics, Tokyo Institute of Technology, Oh-okayama, Meguro-ku, Tokyo 152-8551, Japan}

%\pacs{}
\date{\today}
\begin{abstract}
Using the topological flux insertion procedure, the ground-state
degeneracy of an insulator on a periodic lattice with filling factor
$\nu=p/q$ was found to be at least $q$-fold. Applying the same
argument in a lattice with edges, we show that the degeneracy is modified by the additional edge density $\nu_{\mysw{E}}$ associated with the open boundaries. In particular, we demonstrate that these edge corrections may even make an insulator with integer bulk filling degenerate.
\end{abstract}
\maketitle

A fascinating aspect of quantum mechanics is the interplay between 
global topology and some physical properties, that are naively
believed to depend only on local quantities. Restrictions that originate
from {\it topological} arguments\cite{Wen} often lead to elegant and
non-trivial predictions of bulk properties in a non-perturbative
fashion. One example is particle statistics dictating the possible
ground states of a system. A stark demonstration of that is
anyons with fractional statistics\cite{anyonSC}. Another
instance of this picture is the non-perturbative proof of
Luttinger's theorem\cite{Oshikawa2,Paramekanti04}; it is revealed that the topology dictates the space enclosed within the Fermi surface of a Landau-Fermi liquid to be exactly the particle density in the conduction band, disregarding the details of the microscopic Hamiltonian. Beyond their elegance and appeal, topological ground-state properties have been
recently made the centerpiece, primarily due to Kitaev, of a novel
kind of quantum computation\cite{Kitaev}.

Predictions from topological arguments are often sensitive to the
geometries of the bulk. For instance, the insulating $Z_{2}$-gauge spin
liquid\cite{Senthil00} is four-fold or two-fold degenerate
depending on the bulk geometry is a torus or a cylinder respectively. Another well-known example is the $\nu=1/q$
fractional-quantum-Hall liquid\cite{Wen92} on the two-dimensional
torus. A topological constraint due to these exotic excitations gives
rise to the non-trivial $q$-fold degeneracy. Remarkably, 
by cutting the torus into a closed strip with two open edges, the
ground-state degeneracy in the thermodynamic limit shoots up from $q$
to infinity, with gapless excitations described by the chiral
Luttinger liquid on either side of the Hall bar. It is then natural to
expect that the topological arguments might deliver rather different
messages when edges are present. Indeed, the quantum Hall system is
not the only example of juicy edge physics. Other examples
are Andreev bound states appearing on edges of superconducting
lattices\cite{Hu94}, spin-1/2 excitations of the Haldane phase located
at the edge of $S=1$ Heisenberg chain\cite{AffleckHalperin}, and
the ferromagnetic moment on a zigzag tip of a carbon nanotube\cite{Gil04}.

In this Letter, we revisit a topological constraint in insulators -
the correspondence between the filling factor and the ground state
degeneracy. In Ref. [\onlinecite{Oshikawa1}], Oshikawa extends the
Lieb-Schultz-Mattis argument\cite{Lieb61} and Laughlin's treatment of
the quantized Hall conductance\cite{Laughlin81}, to show that the
ground state degeneracy of an insulator with filling fraction
$\nu=p/q$ (where $q$ and $p$ are coprimes) is at least $q$-fold. This 
is established using the flux-insertion method applied to the
system on the periodic and finite lattice. By taking appropriate
thermodynamic limit, the ground state degeneracy is related to the
intensive particle filling factor in the bulk. However, in the presence of open boundaries, the particle density is no longer
uniform near the edges. The physics of the edge
is captured through appropriate edge filling factors that are
introduced in
addition to the bulk density. In this Letter, we will show that the
degeneracy of the ground state is determined by {\em both the bulk and
edge filling factors} in the presence of edges. 

For instance, a gapped insulator on a periodic lattice, made of spinless
particles with bulk filling factor $\nu_{\mysw{B}}=1$, may have no degeneracy. However, if the 
lattice has open boundary conditions along one
direction, the particle density near the edge may differ from the bulk
filling factor far from the edge. This difference is captured by the
edge density $\nu_{\mysw{E}}$, which arises
from particles localized near the edge, as well as changes of the
extended wave functions in the bulk. The edge state may break translational invariance and gives rise to ground-state degeneracy despite of the bulk state having perfect translational symmetry parallel to the edge. For example, for $\nu_{\mysw{E}}=1/2$, the edge may form a CDW with period two. Such a state is doubly degenerate, although the bulk insulator is featureless. 

Let us now review the flux insertion procedure in the context
of a system with open boundaries in one direction (we follow
Ref.[\onlinecite{Oshikawa1}] closely). Consider a lattice in
$d$-dimensional space with open boundaries in one direction, $1
\le x_1 \le L_1$, and periodic in the remaining $d$-1 dimensions with
finite length $L_{i}$, $i=2,\ldots,d$. The insulating ground
state $|\Phi_{0}\rangle$ contains $N$ particles that reside on the
lattice sites. Making use of the periodicity of the lattice, say
in the $x_{d}$ direction, the ground state can be chosen to be an
eigenstate of the momentum operator $\hat{P}_{d}$, carrying momentum
$P_{d}^{0}$ in the $x_{d}$ direction.

We now introduce a fictitious vector potential $\vec{A}$ that couples
with ``unit charge'' to all particles in the system. Protected by the
gap, one can adiabatically insert a flux quanta $\Phi = h c$
through the ring in the $x_{d}$-direction and map $|\Psi_{0}\rangle$ to
$|\Psi_{0}'\rangle$ of the same energy. Because the flux insertion can
be achieved by the constant vector potential $A_{d} = \Phi/L_{d}$ in
the direction of $x_{d}$, which commutes with $\hat{P}_{d}$, the
momentum remains constant in the whole adiabatic procedure, $
\hat{P}_d|\Psi'_0\rangle=P_{d}^0|\Psi'_0\rangle$.

But the flux insertion changes the Hamiltonian from $\H(\Phi=0)$
to a different topological sector $\H(\Phi=h c)$. Before being able to
compare $|\Psi'_0\rangle$ to $|\Psi_0\rangle$, we need to restore the
Hamiltonian to the same topological sector $\H(\Phi=0)$. This is done
using the following large unitary gauge transformation:
\be
U=\exp\l(\frac{2\pi i}{L_d}\summ_{\vec{x}}x_d\hat{n}_{\vec{x}}\r).
\label{3}
\ee
Now $U\H(h c)U^{-1}=\H(0)$, and therefore $U|\Psi'_0\rangle$ is an
eigenket of the original Hamiltonian. Since our system is a gapped
insulator, and the flux insertion was adiabatic, we expect that
$U|\Psi'_0\rangle$ has the same energy as $|\Psi_0\rangle$ (with
$\H(0)$ in mind). 

The momentum of the newfound ground state can be evaluated straightforwardly,
$
\hat{P}_d
U|\Psi'_0\rangle=(U\hat{P}_d+[\hat{P}_d,U])|\Psi'_0\rangle
=\l(P^0_d+2\pi N/L_d\r)U|\Psi'_0\rangle.
$
So the momentum of $U|\Psi'_0\rangle$ is $P'_d=P^0_d+2\pi N/L_{d}$. If
$N$ and $L_{d}$ are mutually prime, then $U|\Psi'_0\rangle$ and
$|\Psi_0\rangle$ are {\it two} degenerate ground states. To relate the
ground state degeneracy to the filling factor, one need to take
appropriate thermodynamic limit. In a fully periodic system,
Oshikawa's argument follows by saying that 
\be
\Delta P = P'_d-P_d^0=2\pi\frac{N}{L_d}= 2\pi \nu C_{d},
\label{5}
\ee
where $\nu=N/V=p/q$ is the filling factor of the lattice. The volume
of the system is $V=\prod_{i=1}^{d} L_{i}$ and the transverse size at
each particular $x_{i}$ is $C_{i}=V/L_{i}$. The thermodynamic limit is
taken by fixing $\nu$ (so that $p$ and $q$ are well defined without
any correction) and choosing $C_{d}$ to be coprime with
$q$. Thus, by repeating the flux insertion, we can
generate $q$-fold distinct ground states. If we assume that the ground
state degeneracy $D_{g}$ is a robust quantity, disregarding the
details of the thermodynamic limit, the gauge argument leads to the
constraint $D_{g} \ge q$.

Returning to the non-periodic system introduced above, we pick up the
discussion from Eq. (\ref{5}). Two different kinds of non-uniformity
occur near the edges. First, the edges give rise to new states of
particles localized near the edges; we define the {\em edge-state
filling factor} $n_e=N_{\mysw{E}}/C_1$, where $N_{\mysw{E}}$ is the
number of edge particles. Second, bulk states (forming a continuum)
that propagate throughout the lattice may be repelled from the
edges. This repulsion creates a charge depletion near the edge; we
define $n_{\mysw{\Delta}}=N_{\mysw{\Delta}}/C_1$, where
$N_{\mysw{\Delta}}$ is the total charge depleted near the edge. Both
effects give rise to non-trivial corrections to the gauge argument.

Loosely speaking, the presence of edges divides the total number of particles into two groups: bulk and edge, labeled by $N_{\mysw{B}}$ and $N_{\mysw{E}}$ respectively. The precise distinction between bulk and edge states would be explained in later paragraphs. In light of this division, we need to restate Eq. (\ref{5}) as 
\beqn
\Delta P= 2\pi\frac{N_{\mysw{B}}+N_{\mysw{E}}}{L_d}
=2\pi\frac{N_{\mysw{B}}}{L_d}+2\pi n_{e} C_{1d},
\label{6}
\eeqn
with $C_{1d}=V/(L_{1} L_{d})$. Before we can conclude
anything about the ground-state degeneracy, we also need to
account for the repulsion of bulk states from the edge.

As mentioned above, the edges may repel the bulk states, creating a
charge depletion near the edges with density
$n_{\mysw{\Delta}}=N_{\Delta}/C_1$. 
Now, in order to achieve the thermodynamic limit of an insulator with
bulk filling $\nu_B$ 
and edge filling $n_e$, we need to consider a sequence of finite lattices with a total number of particles which is 
\begin{eqnarray}
N_0 &=& N_{\mysw{B}}+N_{\mysw{E}}= (\nu_{\mysw{B}} V -N_{\mysw{\Delta}})+n_e C _1
\nonumber\\
&=& \nu_{\mysw{B}} V+\l(n_e-n_{\mysw{\Delta}}\r)C_1
\label{7}
\end{eqnarray}
Note that $n_{\mysw{\Delta}} \equiv
p_{\mysw{\Delta}}/q_{\mysw{\Delta}}$ is not necessarily an
integer. To constrast, in a periodic system the thermodynamic limit would simply be taken by considering a sequence of lattices with $N_0=\nu_{\mysw{B}} V$
particles. 

Including both types of edge effects, the momentum difference
between $|\Psi_0\rangle$ and its sibling $U|\Psi'_0\rangle$ is 
\be
\Delta P=2\pi \nu_{\mysw{B}} C_{d} + 2\pi \nu_{\mysw{E}} C_{1d}
\label{9},
\ee
with $\nu_{\mysw{B}}=p_{\mysw{B}}/q_{\mysw{B}}$ and $\nu_{\mysw{E}} =
n_e - n_{\mysw{\Delta}} = p_{\mysw{E}}/q_{\mysw{E}}$. From the revised gauge argument in Eq. (\ref{9}), we obtain the main result
of this paper: the degeneracy of an insulating state in the presence of edges is given by
\be
D_{g} \ge LCD\l(q_{\mysw{B}}, q_{\mysw{E}}\r) \ge q_{\mysw{B}},
\label{95}
\ee 
where LCD denotes {\it least common denominator}. 

Before discussing possible extensions and generalizations of the
result in Eq. (\ref{95}), we resolve some subtlties in the derivation above.
A central issue is the difficulty of distinguishing edge
and bulk particles as in Eq. (\ref{6}) in a strongly correlated
system, where the concept of single particle states is rather vague. At first sight, this distinction may seem impossible to carry
out in practice, and might be mistaken to be a theorist's whim; there
is, however, a precise way of counting the number of edge and bulk
states. This method utilizes a spectral function decomposition of the
system's correlation function, which in principle we know with utmost
precision. Following Lehman decomposition of the single-particle
spectral function:
\beqn
A(\x,\o)&=&\sum_m\delta(\o-E_m) |\langle m|\psi^{\dagger}(\x)|\Psi_0\rangle|^{2}
\nonumber\\
&\pm& \sum_m\delta(\o+E_m) |\langle m|\psi(\x)|\Psi_0\rangle|^{2}.
\label{17}
\eeqn
Here $\psi(\x)$ is the annihilation operator in the Schr\"odinger
picture at lattice site $\x$. The $\pm$ distinguishes between fermions
and bosons. Without loss of generality, let us concentrate on fermions.
Note that $|m\rangle$ are many-body excited states that show one particle
excitations. The excitations correspond to one-particle states of free
particles. In the following we simply refer to $|m\rangle$ as states.

Next, consider the $\o<0$ (hole) part of the spectral function
$A(\x,\o)$. In a finite system, it is given as a discrete sum of $\delta$-functions multiplied by matrix elements, and it obeys the sum rule
\be
\summ_{\x}\intt_{-\infty}^{0} A(\x,\o)d\o=N,
\label{18}
\ee
where $N$ is the total number of particles in the lattice. If we take
the limit $L_1\rightarrow \infty$ while keeping the bulk density
fixed, we expect that some of the $\delta$-functions will merge into
continuum, whereas the rest will remain sharp and seperated. 
The first group is the excited {\em bulk states}, and
the second group is the excited {\em edge states}. We stress that
$|m\rangle$ are strongly correlated many-body states that show one particle excitations $\Delta Q = \pm 1$ relative to the ground state. More precisely, bulk states $|m_{\mysw{B}}\rangle$ scale as
\be
|\langle m_{\mysw{B}}|\psi^{\dag}(\x)|\Psi_0\rangle|^{2} \sim {\cal O} \left(
\frac{1}{L_1} \right)
\label{19}
\ee
for any $\x$, as $L_1$ is scaled to infinity. Unlike bulk
states, the edge states, $|m_{\mysw{E}}\rangle$, tend to an $\x$-dependent
constant as $L_1 \to \infty$:
\be
|\langle m_{\mysw{E}}|\psi^{\dag}(\x)|\Psi_0\rangle|^{2} \sim {\cal O}(1).
\label{20}
\ee
Thus we can break the spectral function into a bulk ($A_{\mysw{B}}$) and edge
($A_{\mysw{E}}$) parts:
\beqn
A_{\mysw{B/E}}(\x,\o) &=& 
\sum_{m_{\mysw{B/E}}}\delta(\o-E_{m_{\mysw{B/E}}})
|\langle m_{\mysw{B/E}}|\psi^{\dagger}(\x)|\Psi_0\rangle|^{2}
\nonumber\\
&& \hspace{-1cm} \pm  \sum_{m_{\mysw{B/E}}}\delta(\o+E_{m_{\mysw{B/E}}})
|\langle m_{\mysw{B/E}}|\psi(\x)|\Psi_0\rangle|^{2}.
\label{21}
\eeqn
The definition of $N_{\mysw{B}}$ and $N_{\mysw{E}}$ is hence
\be
N_{\mysw{B(E)}}=\summ_{\x}\intt_{-\infty}^{0}d\o A_{\mysw{B(E)}}(\x,\o).
\label{22}
\ee

There are two subtleties in the above procedure. First, it is possible
that an edge state hybridizes with a bulk state. Definition along
the lines of Eqs. (\ref{21}) and (\ref{22}) specifies that such a
state is an edge state; even though such a state is partially delocalized, it is associated
with the edge due to the localized weight. The number of such states
will scale with $L_i$ for $i>2$, but {\it not}  with $L_1$. Another
subtlety occurs if there is an accidental degeneracy between a bulk
and edge states. In this case, in a finite system, it may be
impossible to distinguish between the edge and bulk states and two
hybridized states. This accidentel degeneracy will surely be lifted by
a different choice of system dimensions, and therefore needs not bother us.

Now that the derivation is complete, we would like to elaborate on some
extensions of the edge argument. The first question we address is the effect of bulk-state depletion near the edge. Note that if $n_{\mysw{\Delta}}$ is an integer, the bound on the ground-state degeneracy, $D_{g} \ge LCD(q_{\mysw{B}},q_{e})$, only depends on $\nu_{\mysw{B}}$ and $n_e=p_{e}/q_{e}$. In this case, the depletion of bulk states does not play any role in determining ground-state degeneracy. In fact, for featureless insulators with low-energy excitations described by Fermi liquid theory, we can show that $n_{\mysw{\Delta}}$ is an integer.

Start with the ground state $|0\rangle$ of the featureless insulator on the periodic lattice (without edges). The insulator with open edge can be viewed as perturbing the original ground state $|0\rangle$ with hole excitations near the boundary. The translational invariance ensures that we can construct local quasi-hole states $|r\rangle = \psi(r) |0\rangle$, with the same spatial profile but located at different $C_{1}$ lattice sites on the edge. In general, these states, located at different lattice sites, would have non-vanishing overlaps. If one assumes the Fermi-liquid picture is at work here, these local (Wannier) orbitals would form a continuous band (with dispersion needs to be determined self-consistently). To ensure the new ground state (in the presence of open boundaries) is also insulating, these quasi-particles must form a {\em band insulator} at the edge, i.e. the number of holes in one unit cell, $n_{\mysw{\Delta}} = N_{\mysw{\Delta}}/C_{1}$, is an integer.

The argument also applies to a simple but exotic situation when the low-energy excitations carry fractional ``charges" (not necessarily the ordinary electric charge), but are well described by Fermi-liquid like theory. Following similar argument, $n_{\mysw{\Delta}}$ is fractional, reflecting the breaking up of ordinary particles. While the argument does not hold when low-energy excitations are no longer described by Fermi liquid theory, in the above case, $n_{\mysw{\Delta}}$ serves as a useful precursor for identifying low-energy excitations in featureless insulators.

Another extension of the theorem is when the edges are far apart. It
is then natural to divide the edge corrections, $n_e = \sum_{i}
n_{ei}$ and $n_{\mysw{\Delta}} = \sum_{i} n_{\mysw{\Delta}i}$,
associating with each edges. A simple example helps to demonstrate this
point and the statements above; consider a segment of carbon nanotube with circumference
$L_{y}$ (in $y$ direction) and zigzag edge at $x=0$ and
$x=L_x$. Choose the tight-binding hopping to be in the strong
anisotropic regime, $|t_{h}|>2|t|$, where $t_{h}$ and $t$
denote the hopping amplitudes along horizontal and (vertical) zigzag
bonds. The interactions between particles are assumed to be weak, with
a typical energy scale $V$.

Ignoring the weak interaction momentarily, the open boundaries at $x=0$ and $x=L_x$, give rise to single-particle edge states at $E=0$ with
momentum-dependent localization length $ \xi(k_y) = \ln |(2t \cos
k_y)/t_{h} |$. Each edge has $L_{y}$ distinct states, corresponding
to each quantized $k_y$. Since the bare hopping Hamiltonian is
quadratic, the total number of states has to equal the number of
particles in the lattice, or twice the number of unit cells (each
carbon-nanotube unit cell contains two sites). Hence, the $L_{y}$ edge
states must come at the expense of $L_{y}$ bulk states. This leads to
$n_{\mysw{\Delta}} =1$ at natural filling $\nu_{\mysw{B}}=1$, and thus
there is no effect on the degeneracy due to the bulk states repulsion
at the edge.

Nevertheless, the edge-state filling factor, $n_e=n_{e1}+n_{e2}$, may be tuned to arbitrary fractions by removing or adding a small
amount of electrons (proportional to the length $L_{y}$) which do not
affect $\nu_{\mysw{B}}$. Because the interaction $V$ is weak (compared
with the bulk gap $\Delta_{\mysw{B}}$), the non-trivial mixing of
single-particle state occurs
within edge states and lifts the exact degeneracy at $E=0$. The ground
state is non-trivial and certainly depends on the specific form of
interactions. But the revised gauge argument with the momentum
shift in Eq. (\ref{9}), leads to at least $q_e$-fold degeneracy of the
ground state even though the bulk filling is an integer.

When considering particular physical systems, as in the above example,
it is possible that the ground-state degeneracy will be larger than
our rigorous result in Eq. (\ref{95}). If the two edges are known by
other means to be independent (as is probably the case in many large
systems) and are also independent of the bulk state, then the degeneracy should be at least the
product $q_{e1}\cdot q_{e2} \cdot LCD (q_{\mysw{\Delta}},
q_{\mysw{B}})$. This degeneracy would become the number of low-lying
states, if there is a weak interaction between the two edges. Other cases are also possible,
most notably, independence of the edge states from the bulk, but not
from each other. In this case the degeneracy is $LCD(q_{e1},\cdot
q_{e2}) \cdot LCD (q_{\mysw{\Delta}}, q_{\mysw{B}})$. We add that for
incommensurate edge filling in the thermodynamic limit $L \to \infty$,
it is very likely that the low-energy physics of the system is
described by a gapless liquid on the edge, living inside the gapped
bulk spectrum (similar to the chiral edge states in
quantum Hall liquid). Since these edge excitations sometimes are the
only low-energy excitations, appropriate treatment of them is
crucially important. 

An interesting situation that may also be quite common is that the bulk
and edge conspire to produce a uniform particle density even
near the edge. This happens if the bulk states repelled from the edge
are exactly compensated by particles trapped at the edge: $n_e =
n_{\mysw{\Delta}}$. Our revised gauge argument is then reduced to
Oshikawa's original argument with only the bulk filling determining
the degeneracy.

The revised gauge argument can be generalized to the case of more than
one non-periodic direction. We require that at least one of the
directions of the lattice is periodic to derive a generalized gauge
argument, although this restriction may not be necessary in a physical system. We consider
$x_d$ as periodic. If all other sides of the attice terminate at
$x_i=0$ and $x_i=L$, there may be edge (surface) and wedge states,
with filling numbers that scale as $N_{\alpha} \sim L^{\alpha}$, where
$\alpha=1,2,...,d-1$. One then needs more filling factors
$n_{\alpha}$ to specify the filling of the system. Also, one needs
the depletion parameters $n_{\mysw{\Delta}a}^b$ where $a=1,\ldots,d$
specifies the repelled `bulk' state, and $b=1,\ldots,a-1$ is the
number of restricted dimensions for the bulk states. To clarify this
statement consider the example of a cube that is made periodic in the
$z$ direction.  The bulk is three dimensional, and the depletion away
from the surfaces ($x=0$, $x=L$, $y=0$ and $y=L$) amounts to
$n_{\mysw{\Delta}3}^1 \cdot L^{2}$ particles. Also, the wedges at
$x=0,y=0$, $x=0,y=L$ etc. may repel the bulk states with a volume
of $n_{\mysw{\Delta}3}^2 \cdot L$. Similarly, edge (surface) states on
the faces of the cube may be repelled from the wedges with corrections
$n_{\mysw{\Delta}2}^{1} \cdot L$.

In summary, using topological flux insertion procedure, we rederive
the connection between the ground-state degeneracy in an insulator with its
filling factors $\nu_{\mysw{B}}$ in the bulk and $\nu_{\mysw{E}}$ at
the edge. As expected, the presence of open edges produces rich physical
phenomena at the boundaries. In addition, the response of the
bulk states to the edge may lend a different perspective into the correlated bulk physics.

We would like to thank Leon Balents, Matthew Fisher, Arun Paramekanti,
Tami Pereg-Barnea, and particularly Masaki Oshikawa for insightful discussions. HHL is grateful for support from NSC-91-2120-M-007-001 and
NSC-92-2112-M-007-039. GR thankfully acknowledges support from  NSF
fund PHY99-07949.


\begin{thebibliography}{10}

\bibitem{Wen} 
X. G. Wen, cond-mat/9506066.

\bibitem{anyonSC}
Y.-H. Chen, F. Wilczek, E. Witten, and B. I. Halperin, 
Int.  J. of Mod. Phys. B, {\bf 3}, 1001 (1989). 

\bibitem{Oshikawa2} 
M. Oshikawa, 
Phys. Rev. Lett. {\bf 84}, 3370 (2000).

\bibitem{Paramekanti04}
A. Paramekanti and A. Vishwanath, cond-mat/0406619.

\bibitem{Kitaev}
A. Yu. Kitaev,
Annals Phys. {\bf 303}, 2 (2003).

\bibitem{Senthil00}
T. Senthil and M.P.A. Fisher, Phys. Rev. B {\bf 62}, 7850 (2000).

\bibitem{Wen92}
W.-G. Wen, Int. J. Mod. Phys. B {\bf 6}, 1711 (1992).

\bibitem{Hu94}
C.-R. Hu, Phys. Rev. Lett. {\bf 72}, 1526 (1994).

\bibitem{AffleckHalperin} 
P. P. Mitra, B. I. Halperin, and I. Affleck, 
Phys. Rev. B, {\bf 45}, 5299 (1992).

\bibitem{Gil04}
G. Refael, H.-H. Lin and M. Oshikawa, work in progress.

\bibitem{Oshikawa1}
M. Oshikawa,
Phys. Rev. Lett. {\bf 84}, 1535 (2000).

\bibitem{Lieb61}
E. H. Lieb, T. Schultz and D. J. Mattis,
Ann. Phys. (N.Y.) {\bf 16}, 407 (1961).

\bibitem{Laughlin81}
R. B. Laughlin,
Phys. Rev. B {\bf 23}, 5632 (1981).

\end{thebibliography}
\end{document}